# Digital Blues: An Investigation into the Use of Bluetooth Protocols

William Brian Ledbetter
*Cybersecurity R&D S&E*
*Cross Domain Assurance*
*Sandia National Laboratories*
Albuquerque, NM, USA
wbledbe@sandia.gov

William Bradley Glisson
*Department of Computer Science*
*College of Science & Eng. Tech.*
*Sam Houston State University*
Huntsville, TX, USA
glisson@shsu.edu

Todd McDonald
*Department of Computer Science*
*School of Computing*
*University of South Alabama*
Mobile, AL, USA
jtmcdonald@southalabama.edu

Todd R. Andel
*Department of Computer Science*
*School of Computing*
*University of South Alabama*
Mobile, AL, USA
tandel@southalabama.edu

George Grispos
*School of Interdisciplinary Informatics*
*College of Information Science & Tech.*
*University of Nebraska at Omaha*
Omaha, NE, USA
ggrispos@unomaha.edu

Kim-Kwang Raymond Choo
*Department of Information Systems &*
*Cyber Security, College of Business*
*University of Texas at San Antonio*
San Antonio, TX, USA
raymond.choo@fulbrightmail.org

*Abstract* - The proliferation of Bluetooth mobile device communications into all aspects of modern society raises security questions by both academicians and practitioners. This environment prompted an investigation into the real-world use of Bluetooth protocols along with an analysis of documented security attacks. The experiment discussed in this paper collected data for one week in a local coffee shop. The data collection took about an hour each day and identified 478 distinct devices. The contribution of this research is two-fold. First, it provides insight into real-world Bluetooth protocols that are being utilized by the general public. Second, it provides foundational research that is necessary for future Bluetooth penetration testing research.

*Keywords-Bluetooth, Bluetooth Detection, Mobile Devices*

## I. INTRODUCTION

Mobile phones are ubiquitous fixtures in today's modern society. Forester predicts that global smartphone subscriptions will hit 3.8 billion and reach 66% of the population by 2022 [1]. Coupling this information with Statista data indicating that worldwide Bluetooth-enabled devices will reach 10 billion by the end of 2018 highlights the idea that Bluetooth technology is not limited to mobile phones. The technology is also pervasive in a range of other device types [2-4].

IHS Technology asserts that "the market for consumer health and fitness electronics is the fastest growing segment of the medical electronics industry" [4]. Informa Telecoms & Media reported that there were over 160 Bluetooth Low Energy fitness devices produced in 2014, encompassing more than 70 distinct device categories [2]. Bluetooth Low Energy is used in fitness wearables from numerous manufacturers, such as Fitbit, Apple, Garmin, Samsung and Xiaomi [5]. Bluetooth-enabled devices are starting to be visible in mainstream healthcare devices. A recent report states that a Bluetooth-enabled heart monitor was recently implanted in a patient [6].

The native support for Bluetooth in all major smartphone and tablet operating systems also makes it an appealing technology for managing smart home accessories utilizing the Internet of Things (IoT) [2, 7]. For the same reason, Bluetooth is becoming a fixture in the automotive industry [2]. Bluetooth is currently used in hands-free communication and is a potential wireless solution for the On-Board Diagnostics (OBD) market [3]. It is starting to be used by smartphone apps to unlock and start automobiles [8]. As Ibn Minar and Tarique [9] note, "One of the major advantages of Bluetooth technology is that it operates in a license-free Industrial, Scientific and Medical (ISM) band ranging from 2.4 to 2.4835 GHz". This makes Bluetooth an attractive interface technology along with presenting possibilities for improving device-to-device communication and capability expansion, much like research into mesh network development [10-12].

The evolution of this environment prompts the idea that Bluetooth capabilities present a logical attack vector and that the technology potentially puts both individuals and businesses at risk. To investigate this idea, an understanding of the practical pervasiveness of various Bluetooth protocols needs to be examined in conjunction with identifying documented viable attacks on Bluetooth-enabled devices.

This research paper is derived from an MSc thesis submission at the University of South Alabama [13]. The contribution of this research is two-fold. First, it provides insight into real-world Bluetooth protocols that are being utilized by the public. It also provides insight into tool capabilities that are freely available for conducting research into detecting Bluetooth protocols. Second, it provides foundational research that is necessary for future Bluetooth penetration research, device vulnerability, and risk mitigation strategies.

The balance of the paper is structured as follows: Section II examines published Bluetooth vulnerability research.

Section III presents the methodology used in this case study. Section IV presents the data, discusses results and explores observations. The last section draws conclusions and presents ideas for future work in this area

## II. LITERATURE REVIEW

The pervasiveness of mobile device residual data [14-16], the security implications associated with mobile devices [17-19], and inherent Bluetooth capabilities in many of these devices is prompting interest from both academicians and practitioners. Coupling this information with research that indicates mobile device residual data, in general, is increasingly being used in a legal context heightens the necessity to understand relevant Bluetooth vulnerabilities [20, 21].

Recent research shows that despite iterative advances and improvements made by the Bluetooth Special Interest Group (SIG), inherent vulnerabilities remain in Bluetooth protocols that can allow devices to be exploited [22]. For example, O'Connor and Reeves [23] noted that Bluetooth is often targeted by attackers because it allows devices to be accessed without a physical connection. One specific vulnerability allows a Man-in-the-Middle (MitM) attack to exploit the pairing process between two devices [24-26]. More recently, researchers have also outlined vulnerabilities associated with the newest Bluetooth protocol, Bluetooth Low Energy (BLE), which is utilized in fitness and medical devices to reduce power consumption and prolong battery life [27, 28]. While many of these vulnerabilities have been reported in the last decade, research has shown that weaknesses in the Bluetooth protocol have existed since its creation [29].

While some researchers have examined the susceptibility of Bluetooth technology, others have focused on vulnerabilities specific to specific protocol versions. Table I presents Bluetooth vulnerabilities that have been documented in peer-reviewed journals and conference papers which specify weaknesses in distinct Bluetooth versions [24-33].

### A. Bluetooth Research

Whenever the Bluetooth core specification is updated, new features are added along with patches for known bugs. Moreover, SIG has also attempted to enhance Bluetooth security standards in response to any vulnerabilities reported in academia [24, 29, 33]. For example, as shown in Table I, Jakobsson and Wetzel [29] revealed the potential for attackers to eavesdrop on Bluetooth connections with a MitM attack. Jakobsson and Wetzel also demonstrated the relative ease with which attackers could circumvent the encryption keys employed in the first Bluetooth core specification due to the limited bit space allotted for encryption keys. Based on these findings, Jakobsson and Wetzel suggested that Bluetooth technology could be used to track devices and predict patterns of human behavior.

In 2003, Kügler [32] extended the work of Jakobsson and Wetzel [29] by demonstrating a similar MitM attack on devices using Bluetooth v1.1. While Jakobsson and Wetzel [29] were only able to initiate a MitM attack if both legitimate devices were set as either master or slave, Kügler [32] was able to do so regardless of the master/slave configuration of the devices. Separately, Hager and Midkiff [30, 31] demonstrated how it was possible for an attacker to successfully spoof a Bluetooth address used by another legitimate device, despite the update to the Bluetooth core specification (from v1.1 to v1.2). The idea behind this attack was that the spoofed address could then be used to initiate a connection to other unsuspecting users as if it were the original device. Hager and Midkiff [30, 31] also confirmed the encryption vulnerability first reported by Jakobsson and Wetzel [29].

In 2007, the Bluetooth SIG released core specification v2.1, which is noted for introducing the Secure Simple Pairing (SSP) protocol to Bluetooth [34]. This was the first major security update that SIG published for Bluetooth, providing a considerable improvement to the pairing process utilized in previous versions [34]. SSP is based upon the Elliptic Curve Diffie-Hellman (ECDH) exchange, in which devices trade public keys and then create a secret key (known as the DHkey) utilized to encrypt further communications [35].

TABLE I. DOCUMENTED VULNERABILITIES

| Version | Release Year | Vulnerability | Citation |
|---|---|---|---|
| 1.0 | 1999 | 1. Security Keys vulnerable to eavesdropping | 1. Jakobsson & Wetzel [29] |
| | | 2. Security Keys vulnerable to cracking | 2. Jakobsson & Wetzel [29] |
| 1.1 | 2002 | 1. Transmission vulnerable to MitM* attack | 1. Kügler [32] |
| 1.2 | 2003 | 1. Device addresses vulnerable to spoofing | 1. Hager & Midkiff [30, 31] |
| 2.0 | 2004 | 1. Security Keys vulnerable to eavesdropping | 1. Shaked & Wool [33] |
| | | 2. Security Keys vulnerable to cracking | 2. Shaked & Wool [33] |
| 2.1 | 2007 | 1. SSP PE** vulnerable to MitM* attack | 1. Barnickel, et al. [26] |
| | | 2. SSP PE** vulnerable to password guessing | 2. Lindell [24] |
| 3.0 | 2009 | - Specific vulnerabilities not yet found | N/A |
| 4.0 | 2010 | 1. SSP PE** still vulnerable to MitM* attack | 1. Phan & Mingard [25] |
| | | 2. BLE*** vulnerable to passive eavesdropping | 2. Ryan [27] |
| | | 3. BLE*** vulnerable to MitM* attack | 3. Xu, et al. [28] |
| 4.1 | 2013 | - Specific vulnerabilities not yet found | N/A |
| 4.2 | 2014 | - Specific vulnerabilities not yet found | N/A |

*MitM = Man in the Middle
**SSP PE = Secure Simple Pairing protocol in Passkey Entry mode
***BLE = Bluetooth Low Energy

While in theory, SSP is a secure protocol, various researchers identified that it too contains vulnerabilities. [24, 26]. By exploiting the use of an unencrypted connection for the creation of the secret key, Lindell [24] mounted successful attacks that exposed the weakness in the SSP process between

Bluetooth devices. Barnickel et al. [26] extended Lindell's [24] concerns discovered in the SSP protocol by taking advantage of PassKey Entry mode transmissions conducted in the clear.

*B. Bluetooth Low Energy Research*

In 2010, the Bluetooth SIG released core specification v4.0, which introduced Bluetooth Low Energy (BLE) to promote a focus on devices with less power consumption [36]. While previous versions of Bluetooth were limited to larger devices such as headphones, laptops, and mobile phones, the energy conservation of Bluetooth Low Energy allowed it to be utilized in much smaller devices powered by coin-cell batteries [13]. Core versions 4.1 and 4.2 (released in 2013 and 2014, respectively) expanded the functionality of BLE by incorporating compatibility with the Internet of Things (IoT) and utilizing the IP protocol to decrease power consumption further [13].

Since Bluetooth v2.1 introduced the Secure Simple Pairing (SSP) protocol in 2007, five core specification updates have been released; v3.0 in 2009, v4.0 in 2010, v4.1 in 2013, v4.2 in 2014, and v5.0 in December 2016 [37, 38]. While there have been several releases, no documentation has been found to indicate a core specification issued by the Bluetooth SIG properly addresses the specific security issues that have been identified in the literature [24, 26, 39]. The continued use of SSP in the Bluetooth protocol stack allows nearly all Bluetooth-enabled devices to remain vulnerable to MitM attacks, regardless of manufacturer or Bluetooth protocol version [40]. Phan and Mingard [25] demonstrated that the SSP protocol used in Bluetooth v4.0 (released 2010) is just as vulnerable to MitM attacks as it is in v2.1 (released 2007). These researchers were able to successfully replicate the attack methods utilized by Lindell [24] four years later. This weakness remains because SSP in Bluetooth v4.0 is implemented in the same flawed way as it originally was implemented in v2.1 [34, 41]. No significant updates have been issued by the Bluetooth SIG to patch this known vulnerability [42]. Phan and Mingard further confirmed Lindell's original findings; SSP Passkey Entry mode remains the most utilized Bluetooth authentication method to date, prompting it to be the most vulnerable to attack [24, 25].

While Bluetooth Low Energy is much more energy efficient than its Bluetooth Classic predecessor, research shows that these sacrifices in power and sophistication do not come without a cost [43]. Xu et al. [28] determine that BLE devices are susceptible to MitM attacks because of the vulnerabilities by incorporating Bluetooth Low Energy into the Bluetooth paring protocol. Similarly, Ryan [27] postulates that the concessions made to streamline the Bluetooth Low Energy core specification weaken the security of the protocol and make data transmitted over BLE connections vulnerable to simple attack techniques. The researcher goes on to demonstrate how readily-available open source software and hardware can be utilized to passively eavesdrop on BLE transmissions and to even break its encryption scheme with little effort. Ryan concludes that these design choices essentially remove any privacy in the Bluetooth Low Energy protocol.

Previous research indicates that vulnerabilities are present in a variety of Bluetooth protocols that have been introduced over the years. The question that persists is to what extent are these protocols being used in a real-world context.

III. METHODOLOGY

Despite increasing security concerns, Bluetooth users continue to use vulnerable versions of the technology's protocol. To acquire an understanding of the versions of Bluetooth that are currently being used in the wild, a case study was implemented using a design science approach. Hevner et al. [44] define Design Science Research (DSR) as the creation of technological artifacts to provide solutions for problems that exist in the real world. Hevner, et al. [44] provide a framework for conducting DSR which includes the phases: (1) identification of the research problem and motivation for finding a solution; (2) definition of solution objectives; (3) design and development of the artifact; (4) demonstration of the solution; (5) evaluation of the solution; (6) communication of the research problem and solution.

The research problem presented in the introduction focused on the construction of a tool to identify Bluetooth versions and devices that are used in-the-wild. The solution objective is to capture as much information that is being broadcast by a device as possible. The design and development involved the identification of the Ubertooth One antenna, and viable open source software that could be downloaded, i.e., Blue Hydra. Blue Hydra runs on Linux and is compatible with Ubuntu. Blue Hydra utilizes a BlueZ Linux library to communicate with Bluetooth. The demonstration of the software and tool configuration was initially tested in a lab to be sure that it was working. The iterative process resulted in the development of a toolset that could be utilized to collect transmitted Bluetooth data. The components of the toolkit developed in this research are available in Table II. A screenshot of the tool is provided in Figure. 1. The evaluation of the solution is the data collection discussed in the next section.

*A. Data Collection*

Data collection lasted for a week starting on Sunday, December 4, 2016, at a local coffee shop in Mobile, Alabama, USA. The first data collection session lasted for one hour. The time span was increased to ninety minutes the following session, and this persisted as the constant duration for each session for the remainder of the data collection process. There was a concern that the vast amount of data collected by the Blue Hydra software in such a short period could potentially create a memory overflow, resulting in slowed application response, hardware errors, or even a system crash. To account for this possibility, each data collection session was divided into three separate thirty-minute segments. The Blue Hydra software was stopped and restarted (along with the reinitialization of the Ubertooth One device) for each of these segments. For the data collection, the sessions were conducted during the afternoon hours. Start times for each session ranged between 3:15 PM and 4:30 PM Central Standard Time (CST).

TABLE II.  TOOLSET COMPONENTS

| Hardware & Software | Version |
|---|---|
| Toshiba Satellite P775-S7160 Laptop | • Intel Core i7-2670QM CPU @ 2.20GHz x8 processor<br>• Mobile Intel HM65 Express (6MB L3 cache) chipset<br>• 8GB DDR3 SDRAM @1333MHz (2x4GB) memory<br>• 750GB SATA HDD (5400RPM) hard drive<br>• Ubuntu 16.04.1 LTS (64-bit) OS |
| Ubertooth One | Firmware release: 2015-10-R1 |
| SENA Parani-UD100 Bluetooth Adapter | UD100-G03, Class 1 device, BT v4.0 |
| Blue Hydra | https://github.com/pwnieexpress/blue_hydra |
| Ruby | v2.3.1p112 (2016-04-26 revision 54768) [x86_64-linux] |
| DB Browser for SQLite | v3.7.0 (QT v5.5.1, SQLite v3.9.2) |

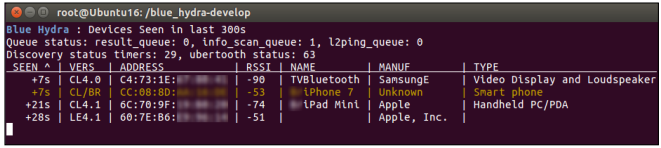

Figure 1. Bluetooth Sniffer

The amount of data allocated for the backup file created during each data collection segment was saved in the data collection log file. Along with the statistical information tallied for the parameters of the data collection process, a total file size accumulation was calculated in the log file to prevent data storage capacity from becoming a prohibitive factor.

### B. Data Storage

The information collected about each device included: the address, name, vendor, chipset manufacturer, as well as the type and version of Bluetooth protocol. Timestamp information was also recorded related to device discovery. A summary of the database columns is presented in Table III.

## IV. RESULTS AND ANALYSIS

The total number of new devices detected for the week from December 4, 2016 to December 10, 2016 was 478. The daily breakdown for device detection is available in Table IV. A range of Bluetooth versions was identified during this research. Two interesting results that need to be investigated in the future include the 103 devices that were discovered running v1.0b and the 189 devices that contained a null value. The number of devices discovered by a version of Bluetooth is presented in Table V. An illustration of the version distribution is offered in Figure 3.

TABLE III.  DATABASE INFORMATION

| Database Column | Description |
|---|---|
| Name | Device name (manufacturer default, if not defined by the user) |
| Address | Bluetooth Device Address (BD_ADDR) |
| Company | Device manufacturer |
| LMP Version | Identifies Bluetooth (BT) version and subversion |
| Manufacturer | Bluetooth chipset manufacturer |
| Classic Mode | Indicates if Bluetooth type identified as BT Classic (T or F) |
| LE Mode | Indicates if Bluetooth type identified as BLE (T or F) |
| Classic RSSI | Timestamps and signal strength values for BT Classic devices |
| LE RSSI | Timestamps and signal strength values for BLE devices |
| Created At | Indicates first time device detected (timestamp) |
| Last Seen | Indicates last time device detected (timestamp) |

TABLE IV.  NEW DEVICES DETECTED DAILY

| Scan Date | New Devices |
|---|---|
| December 4, 2016 | 102 |
| December 5, 2016 | 15 |
| December 6, 2016 | 51 |
| December 7, 2016 | 76 |
| December 8, 2016 | 88 |
| December 9, 2016 | 82 |
| December 10, 2016 | 64 |
| **Total** | **478** |

The average length of time for device detection for the week was 7.75 minutes. The average device detection times for each day are presented in Figure 2.

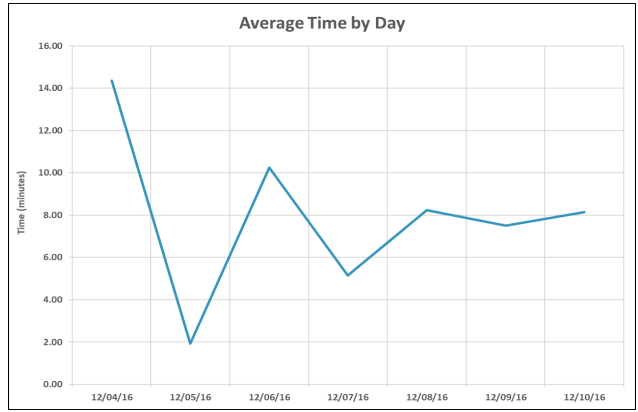

Figure 2. Average Device Detection Time by Day

TABLE V. BLUETOOTH VERSION COUNTS

| Bluetooth Version | Count |
| --- | --- |
| v1.0b | 103 |
| v1.1 | 0 |
| v1.2 | 0 |
| v2.0 | 1 |
| v2.1 | 2 |
| v3.0 | 1 |
| v4.0 | 16 |
| v4.1 | 18 |
| v4.2 | 148 |
| NULL | 189 |
| **Total** | **478** |

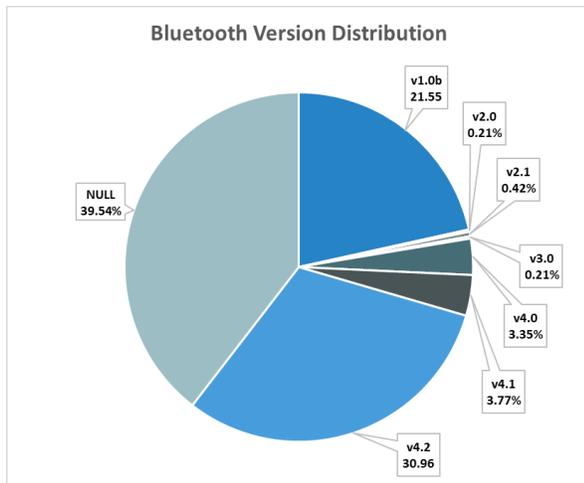

Figure 3. Distribution of Bluetooth Versioning Data

Seventy-five (75) Bluetooth Classic devices and 403 BLE devices were detected. This information is illustrated in Figure 4.

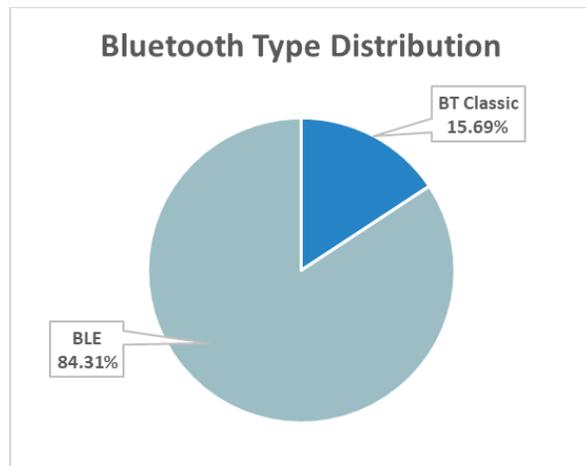

Figure 4. Distribution of Bluetooth Versioning Data

## V. CONCLUSIONS AND FUTURE WORK

The literature review indicates that security vulnerabilities are present in Bluetooth protocols. The data collected by the Bluetooth sniffer developed as part of this case study indicates that a range of Bluetooth protocols is being used in a real-world context. These devices include mobile phones, tablets, computers, televisions, fitness wearables, smart watches, automobiles, headphones, speakers, and location trackers, among others. The initial data collection supports the idea that Bluetooth capabilities present a valid exploitable attack vector that potentially places individuals and organizations at risk.

Future work will build on the database generated by the toolkit developed in this research. The database information can be utilized to determine the viability of launching penetration testing activates. The vitality is determined based on the version information, calculating the distance to the device, and average device range times. Future research will forensically investigate the creation of residual data on a compromised Bluetooth-enabled device, and if industry-accepted mobile device forensic toolkits can detect this data in forensically-sound environments.